\documentclass[aps,prl,twocolumn,showpacs,superscriptaddress,preprintnumbers,floatfix]{revtex4}
\usepackage{graphicx}
\usepackage{xspace}
\usepackage{epsfig}
\usepackage{multirow}
\usepackage{dcolumn}

\newcommand{\de}{\ensuremath{\Delta E}\xspace}

\newcommand{\bb}{\ensuremath{B \overline{B}}\xspace}

\def\myspecial#1{}                   
\def\calL{{\mathcal L}}

\def\Mbc{M_{\rm bc}}

\begin{document}
\preprint{
  Belle Preprint 2012-22
}
\preprint{
  KEK Preprint 2012-19
}

\myspecial{!userdict begin /bop-hook{gsave 300 50 translate 5 rotate
    /Times-Roman findfont 18 scalefont setfont
    0 0 moveto 0.70 setgray
    (\mySpecialText)
    show grestore}def end}

\title{\quad\\[0.5cm]
Measurements of Branching Fractions and Direct $CP$ Asymmetries for 
$B \to K \pi, B\to \pi 
\pi$  and $B\to KK$ Decays} 
%%% Paper:    B0 -> h h
%%% Journal:  Physical Review Letters
%%% Contacts: Y.T. Duh (kaliduh@hep1.phys.ntu.edu.tw)
%%%           T.Y. Wu (r97222020@ntu.edu.tw)
%%%           P. Chang (pchang@hep1.phys.ntu.edu.tw)
%%%           G.B. Mohanty (gmohanty@tifr.res.in)
%%%           Y. Unno (yunno@post.kek.jp)
%%% Non-responding authors or those who said NO are commented out.
%%% ====================================================================
%%% Click the RELOAD button on your web browser to see the updated file.
%%% ====================================================================
%%% Use \input{author} to insert this material into your latex file.
%%%%% Force institutions to appear in alphabetical order when typeset.
%%%\affiliation{University of the Basque Country UPV/EHU, Bilbao}
%%%\affiliation{University of Bonn, Bonn}
\affiliation{Budker Institute of Nuclear Physics SB RAS and Novosibirsk State University, Novosibirsk 630090}
\affiliation{Faculty of Mathematics and Physics, Charles University, Prague}
%%%\affiliation{Chiba University, Chiba}
\affiliation{University of Cincinnati, Cincinnati, Ohio 45221}
%%%\affiliation{Department of Physics, Fu Jen Catholic University, Taipei}
%%%\affiliation{Justus-Liebig-Universit\"at Gie\ss{}en, Gie\ss{}en}
\affiliation{Gifu University, Gifu}
\affiliation{II. Physikalisches Institut, Georg-August-Universit\"at G\"ottingen, G\"ottingen}
%%%\affiliation{The Graduate University for Advanced Studies, Hayama}
%%%\affiliation{Gyeongsang National University, Chinju}
\affiliation{Hanyang University, Seoul}
\affiliation{University of Hawaii, Honolulu, Hawaii 96822}
\affiliation{High Energy Accelerator Research Organization (KEK), Tsukuba}
%%%\affiliation{Hiroshima Institute of Technology, Hiroshima}
%%%\affiliation{IKERBASQUE, Bilbao}
%%%\affiliation{University of Illinois at Urbana-Champaign, Urbana, Illinois 61801}
\affiliation{Indian Institute of Technology Guwahati, Guwahati}
\affiliation{Indian Institute of Technology Madras, Madras}
%%%\affiliation{Indiana University, Bloomington, Indiana 47408}
%%%\affiliation{Institute of High Energy Physics, Chinese Academy of Sciences, Beijing}
\affiliation{Institute of High Energy Physics, Vienna}
\affiliation{Institute of High Energy Physics, Protvino}
%%%\affiliation{Institute of Mathematical Sciences, Chennai}
%%%\affiliation{INFN - Sezione di Torino, Torino}
\affiliation{Institute for Theoretical and Experimental Physics, Moscow}
\affiliation{J. Stefan Institute, Ljubljana}
\affiliation{Kanagawa University, Yokohama}
\affiliation{Institut f\"ur Experimentelle Kernphysik, Karlsruher Institut f\"ur Technologie, Karlsruhe}
\affiliation{Korea Institute of Science and Technology Information, Daejeon}
\affiliation{Korea University, Seoul}
%%%\affiliation{Kyoto University, Kyoto}
\affiliation{Kyungpook National University, Taegu}
\affiliation{\'Ecole Polytechnique F\'ed\'erale de Lausanne (EPFL), Lausanne}
\affiliation{Faculty of Mathematics and Physics, University of Ljubljana, Ljubljana}
\affiliation{Luther College, Decorah, Iowa 52101}
\affiliation{University of Maribor, Maribor}
%%%\affiliation{Max-Planck-Institut f\"ur Physik, M\"unchen}
\affiliation{University of Melbourne, School of Physics, Victoria 3010}
\affiliation{Graduate School of Science, Nagoya University, Nagoya}
\affiliation{Kobayashi-Maskawa Institute, Nagoya University, Nagoya}
%%%\affiliation{Nara University of Education, Nara}
\affiliation{Nara Women's University, Nara}
\affiliation{National Central University, Chung-li}
\affiliation{National United University, Miao Li}
\affiliation{Department of Physics, National Taiwan University, Taipei}
\affiliation{H. Niewodniczanski Institute of Nuclear Physics, Krakow}
\affiliation{Nippon Dental University, Niigata}
\affiliation{Niigata University, Niigata}
%%%\affiliation{University of Nova Gorica, Nova Gorica}
\affiliation{Osaka City University, Osaka}
%%%\affiliation{Osaka University, Osaka}
\affiliation{Pacific Northwest National Laboratory, Richland, Washington 99352}
\affiliation{Panjab University, Chandigarh}
%%%\affiliation{Peking University, Beijing}
%%%\affiliation{Princeton University, Princeton, New Jersey 08544}
\affiliation{Research Center for Electron Photon Science, Tohoku University, Sendai}
%%%\affiliation{Research Center for Nuclear Physics, Osaka University, Osaka}
%%%\affiliation{RIKEN BNL Research Center, Upton, New York 11973}
%%%\affiliation{Saga University, Saga}
\affiliation{University of Science and Technology of China, Hefei}
\affiliation{Seoul National University, Seoul}
%%%\affiliation{Shinshu University, Nagano}
\affiliation{Sungkyunkwan University, Suwon}
\affiliation{School of Physics, University of Sydney, NSW 2006}
\affiliation{Tata Institute of Fundamental Research, Mumbai}
%%%\affiliation{Excellence Cluster Universe, Technische Universit\"at M\"unchen, Garching}
\affiliation{Toho University, Funabashi}
\affiliation{Tohoku Gakuin University, Tagajo}
\affiliation{Tohoku University, Sendai}
\affiliation{Department of Physics, University of Tokyo, Tokyo}
\affiliation{Tokyo Institute of Technology, Tokyo}
\affiliation{Tokyo Metropolitan University, Tokyo}
\affiliation{Tokyo University of Agriculture and Technology, Tokyo}
%%%\affiliation{Toyama National College of Maritime Technology, Toyama}
\affiliation{CNP, Virginia Polytechnic Institute and State University, Blacksburg, Virginia 24061}
\affiliation{Wayne State University, Detroit, Michigan 48202}
%%%\affiliation{Yamagata University, Yamagata}
\affiliation{Yonsei University, Seoul}
  \author{Y.-T.~Duh}\affiliation{Department of Physics, National Taiwan University, Taipei} % Taiwan
  \author{T.-Y.~Wu}\affiliation{Department of Physics, National Taiwan University, Taipei} % Taiwan
  \author{P.~Chang}\affiliation{Department of Physics, National Taiwan University, Taipei} % Taiwan
  \author{G.~B.~Mohanty}\affiliation{Tata Institute of Fundamental Research, Mumbai} % Tata
  \author{Y.~Unno}\affiliation{Hanyang University, Seoul} % Hanyang
  \author{I.~Adachi}\affiliation{High Energy Accelerator Research Organization (KEK), Tsukuba} % KEK
% \author{K.~Adamczyk}\affiliation{H. Niewodniczanski Institute of Nuclear Physics, Krakow} % Krakow
 \author{H.~Aihara}\affiliation{Department of Physics, University of Tokyo, Tokyo} % Tokyo
% \author{K.~Arinstein}\affiliation{Budker Institute of Nuclear Physics SB RAS and Novosibirsk State University, Novosibirsk 630090} % BINP
% \author{Y.~Arita}\affiliation{Graduate School of Science, Nagoya University, Nagoya} % Nagoya
  \author{D.~M.~Asner}\affiliation{Pacific Northwest National Laboratory, Richland, Washington 99352} % PNNL
% \author{T.~Aso}\affiliation{Toyama National College of Maritime Technology, Toyama} % Toyama
  \author{V.~Aulchenko}\affiliation{Budker Institute of Nuclear Physics SB RAS and Novosibirsk State University, Novosibirsk 630090} % BINP
  \author{T.~Aushev}\affiliation{Institute for Theoretical and Experimental Physics, Moscow} % ITEP
% ...
% Karlsruhe
% \author{N.~Zwahlen}\affiliation{\'Ecole Polytechnique F\'ed\'erale de Lausanne (EPFL), Lausanne} % Lausanne
% \author{O.~Zyukova}\affiliation{Budker Institute of Nuclear Physics SB RAS and Novosibirsk State University, Novosibirsk 630090} % BINP
\collaboration{The Belle Collaboration}

\noaffiliation

\begin{abstract}
We report measurements of the branching fractions and direct 
$CP$ asymmetries (${\cal A}_{CP}$)
for $B \to K\pi, \pi\pi$ and $KK$ decays (but not $\pi^0\pi^0$) 
 based on the final data sample of $772\times 10^6$
$\bb$ pairs collected at the $\Upsilon(4S)$ resonance with the Belle
detector at the KEKB asymmetric-energy $e^+ e^-$ collider. 
We set a 90\% confidence-level upper limit for $K^+ K^-$ at $2.0\times 10^{-7}$;
all other decays are observed with 
branching fractions ranging from $10^{-6}$ to $10^{-5}$. 
In the $B^0/{\overline B}{}^0 \to K^\pm\pi^\mp$ mode, 
we confirm Belle's previously reported large ${\cal A}_{CP}$ with a value of
$-0.069\pm 0.014\pm 0.007$ and a significance of $4.4 \sigma$. 
For all other flavor-specific modes, we find 
${\cal A}_{CP}$ values consistent with zero, including ${\cal A}_{CP}(K^+ \pi^0) = +0.043\pm 0.024\pm 0.007$ 
with $1.8 \sigma$ significance.
The difference of $CP$ asymmetry between $B^\pm\to K^\pm\pi^0$ and 
$B^0/{\overline B}{}^0\to K^\pm\pi^\mp$ is found to be $\Delta {\cal A}_{K\pi}\equiv {\cal A}_{CP}(K^+\pi^0) -
{\cal A}_{CP}(K^+\pi^-) = +0.112\pm 0.027\pm 0.007$
 with $4.0\sigma$ significance. 
We also calculate the ratios of partial widths for the $B \to K\pi$ decays.
Using our results, we test the validity of the sum rule
${\cal A}_{CP}(K^+\pi^-) + {\cal A}_{CP}(K^0\pi^+) \frac{\Gamma(K^0\pi^+)}{\Gamma(K^+\pi^-)}
- {\cal A}_{CP}(K^+\pi^0)\frac{2\Gamma(K^+\pi^0)}{\Gamma(K^+\pi^-)} - 
{\cal A}_{CP}(K^0\pi^0)\frac{2\Gamma(K^0\pi^0)}{\Gamma(K^+\pi^-)}=0$ 
and obtain a sum of $-0.270\pm 0.132\pm 0.060$ 
with $1.9\sigma$ significance.
\end{abstract}

\pacs{11.30.Er, 12.15.Hh, 13.25.Hw, 14.40.Nd}
\maketitle

\tighten

{\renewcommand{\thefootnote}{\fnsymbol{footnote}} 
\setcounter{footnote}{0}

Charmless $B$ meson decays to $K\pi, \pi\pi$ and $KK$ final states provide a 
good test bed to understand $B$ decay mechanisms and to search for physics 
beyond the Standard Model (SM). 
Although predictions for the branching fractions under various theoretical approaches 
suffer from large hadronic uncertainties, direct $CP$ asymmetries
and ratios of branching fractions can still provide excellent sensitivity to new physics (NP), 
since many theoretical and experimental uncertainties cancel out in these quantities.
The direct $CP$ asymmetry is defined as 
\begin{eqnarray}
\label{dfacp}
{\cal A}_{CP}\equiv \frac{N(\overline{B}\to \overline{f}) - N(B\to f)} {N(\overline{B}\to \overline{f})+N(B\to f)}
\end{eqnarray}
where $f/\overline{f}$ denotes
a specific final state from a $B^+/B^-$ or $B^0/{\overline B}{}^0$ decay.
For instance, the observed ${\cal A}_{CP}$ difference between $B^\pm \to 
K^\pm \pi^0$ and $B^0/{\overline B}{}^0\to K^\pm \pi^\mp$ \cite{belleacpkpi,babaracphh,babaracpkpi0}, also 
known as the $\Delta {\cal A}_{K\pi}$ puzzle, can be explained by an enhanced 
color-suppressed tree contribution~\cite{sptree} or NP in the electroweak penguin loop~\cite{nptheory}.                
Other variables sensitive to electroweak penguin contributions are the ratios of 
partial widths, e.g., $R_c\equiv 2\Gamma(B^+ \to K^+\pi^0)/\Gamma(B^+ \to K^0\pi^+)$ and 
$R_n\equiv \Gamma(B^0 \to K^+\pi^-)/2\Gamma(B^0 \to K^0\pi^0)$. 
Prior measurements~\cite{babaracpkpi0,bellebr,babarbr,cleobr} 
of these ratios are consistent with theory expectations~\cite{buras,pqcd,rn1,rn2}, albeit with large errors. 
The experimental uncertainties, therefore, need to be 
improved to adequately compare data and SM predictions.

In this paper, we report measurements of the branching fractions for
$B \to K\pi, \pi\pi$, and $KK$ decays, other than $B^0\to \pi^0\pi^0$, 
and of the direct $CP$ asymmetries for the modes with flavor-specific final states~\cite{conju}.
The measurements are based on $772\times10^6$ $\bb$ pairs, corresponding to the final $\Upsilon(4S)$ data set collected with the Belle detector~\cite{aba} at the KEKB $e^+e^-$ asymmetric-energy 
collider~\cite{kur}.
Compared to our previous publications~\cite{belleacpkpi, bellebr, acpk0pi0}, 
we have increased the $KK$, $K^0\pi^+$ and $\pi^+\pi^-$ data samples by about 
72\%, the $K^+\pi^-$, $K^+\pi^0$ and $\pi^+\pi^0$ samples by about 44\%, and 
the $K^0 \pi^0$ sample by about 18\%, 
have included several improvements in reconstruction
algorithms that enhance the reconstruction efficiency for the charged tracks, 
and have made numerous modifications
to the analysis to improve the measurement sensitivity [e.g., by including an
extra discriminating variable in the likelihood fit; see Eq. (\ref{fit})].

\begin{table*}[th]
\begin{center}
\caption{Signal yields, product of efficiencies ($\varepsilon$) and sub-decay
branching fractions $({\cal B}_s)$ \cite{subeff}, measured branching fractions ($\mathcal{B}$), direct $CP$ asymmetries (${\cal A}_{CP}$) after the correction,
and significance of $CP$ asymmetries ($\mathcal{S}$) for 
individual modes. The first and second quoted errors are 
statistical and systematic, respectively. Upper limit is given at the 90\% confidence level.}
\begin{tabular}{lccccccc}
\hline\hline
~Mode~ & Yield & $\varepsilon \times {\cal B}_s$(\%) & ${\cal B}~(10^{-6})$ & ${\cal A}_{CP}$ & &
$\mathcal{S}~(\sigma)$ \\
\hline
~$K^+\pi^-$ &$ 7525\pm 127$ & 48.82 &
$20.00 \pm 0.34 \pm 0.60$ & $-0.069\pm 0.014\pm 0.007$ & & 4.4 \\
~$\pi^+ \pi^-$ & $2111 \pm 89$  & 54.79 & 5.04 $\pm 0.21 \pm 0.18$ & $-$ & & $-$ \\
~$K^+ \pi^0$ &$3731\pm 92$ & $38.30$ &$12.62 \pm 0.31 \pm 0.56$ & $+0.043\pm
0.024\pm 0.002$ & & $1.8$ \\
~$\pi^+ \pi^0$ & $1846\pm 82$ & 40.80 & $5.86\pm 0.26 \pm 0.38$ & $+0.025\pm 0.043\pm 0.007$ & & $0.6$  \\
~${\overline K}{}^0 K^+$ & $134\pm 23$ & 15.64 & $1.11\pm 0.19 \pm 0.05$ & $+0.014\pm 0.168\pm 0.002$ & & $0.1$  \\
~$K^0 \pi^+$ & $3229\pm 71$ & 17.46 & $23.97\pm 0.53 \pm 0.71$ & $-0.011\pm 0.021\pm 0.006$ & & $0.5$ \\
~$K^0 {\overline K}{}^0$ &$103\pm 15$ & 10.61 & $1.26\pm 0.19\pm 0.05$ & $-$ & & $-$  \\
~$K^0 \pi^0$ &$961\pm 45$ & 12.86 & $9.68\pm 0.46\pm 0.50$ & $-$ & & $-$  \\
~$K^+ K^-$ &$35\pm 29$ & $47.72$ & $0.10\pm 0.08\pm 0.04~(<0.20)$ & $-$ & & $-$ \\
\hline\hline
\end{tabular}
\label{tab:br}
\end{center}
\end{table*}

We define our event selection criteria for these measurements as follows.
Charged tracks originating from a $B$ decay
are required to have a distance of closest approach with respect
to the interaction point less than $4.0\,{\rm cm}$ along the beam
direction ($z$-axis) and less than $0.3\,{\rm cm}$ in the transverse plane.
Charged kaons and pions are identified with
information from particle identification detectors,
which are combined to form 
a $K$-$\pi$ likelihood ratio ${\cal R}_{K/\pi}
= \mathcal{L}_K/(\mathcal{L}_K+\mathcal{L}_\pi)$, where
$\mathcal{L}_{K}$ $(\mathcal{L}_{\pi})$ is the likelihood of the
track being a kaon (pion).
Track candidates with ${\cal R}_{K/\pi}>0.6$ $(<0.4)$ are
classified as kaons (pions). The typical kaon (pion) identification  
efficiency is 83\% (88\%)
with a pion (kaon) misidentification probability of 7\% (11\%).
A tighter ${\cal R}_{K/\pi}$ requirement ($>0.7$) is applied for 
the ${\overline K}{}^0K^+$ channel to reduce the $B^+\to K^0\pi^+$ feed-across 
since the ${\overline K}{}^0K^+$ 
branching fraction is an order of magnitude smaller than that of $K^0\pi^+$.
Charged tracks found to be consistent with an electron or a muon are rejected. 
Candidate $K^0$ mesons are reconstructed via $K^0_S\to \pi^+ \pi^-$~\cite{ksdis}
by requiring the invariant mass of the pion pair
to be $480\,{\rm MeV}/c^2<M_{\pi \pi}<516\,{\rm MeV}/c^2$ 
(corresponding to $5.2\sigma$ around the mean value).
Pairs of photons with invariant masses lying in the range of 
$115\,{\rm MeV}/c^2<M_{\gamma \gamma}<152\,{\rm MeV}/c^2$ 
(corresponding to $2.5\sigma$ around the mean value) 
are classified as $\pi^0$ candidates. The photon energy
is required to be greater than 50 (100) MeV in 
the barrel (endcap) calorimeter.

Candidate $B$ mesons are identified using the beam-energy-constrained mass,
$M_{\rm bc} \equiv
\sqrt{E^{*2}_{\mbox{\scriptsize beam}}/c^4 - |{\vec{p}_{B}}^*/c|^2}$, and the energy difference,
$\Delta E \equiv E_B^* - E^*_{\mbox{\scriptsize beam}}$, where
$E^*_{\mbox{\scriptsize beam}}$ is the run-dependent beam energy, and $E^*_B$ 
and $p^*_B$ are the reconstructed energy and momentum of $B$ 
candidates in the center-of-mass (CM) frame, respectively.
Events with $M_{\rm bc}>5.2\,{\rm GeV}/c^2$ and $|\Delta E|<0.3\,{\rm GeV}$
are retained for further analysis.
For decays having a
$\pi^0$ in the final state, the correlation between $M_{\rm bc}$ and 
$\Delta E$ is relatively large due to photon shower leakage in the calorimeter. 
To reduce this correlation, $M_{\rm bc}$ is calculated by scaling 
the measured $\pi^0$ momentum to the value expected for signal, 
given by ${\vec{p}_{\pi^0}}^* = \frac{{{\vec{p}_{\pi^0}}^*}} {|{\vec{p}_{\pi^0}}^*|} 
\sqrt{(E^*_{\mbox{\scriptsize beam}} - E^*_{h^\pm})^2/c^2-m_{\pi^0}^{*2}c^2}$,
where $h^\pm$ represents the charged kaon or pion. Consequently, the 
correlation coefficient falls from $+$18\% to $-$4\%, as shown by a Monte Carlo (MC) study.  

The dominant background arises from $e^+e^- \to q\overline q ~( q=u,d,s,c )$ 
continuum events. We use event topology to distinguish spherical $B\overline{B}$
events from the jet-like continuum background. A set of modified 
Fox-Wolfram moments~\cite{pi0pi02} is combined into a
Fisher discriminant. Signal and background likelihoods are formed 
based on MC events.
The likelihood, $\cal L$, is the product of the probability density 
functions (PDFs) for the Fisher discriminant, the cosine of the 
polar angle of the $B$-meson flight direction in the CM frame, and the flight-length difference ($\Delta z$)
along the $z$-axis
between the decay vertex of the signal $B$ and  
the vertex formed from the tracks not associated with the signal $B$. 
The decay vertices for $B^+ \to h^+ h^0$ (where $h^0$ represents 
$\pi^0$ or $K^0$) candidates are estimated by the point of closest approach 
of the $h^+$ trajectory to  the
$z$-axis. Since the $K^0\pi^0$ mode has no primary charged track, the $\Delta z$ variable is not used. 
A loose continuum suppression requirement of $\mathcal{R}>0.2$  
 rejects more than 70\% of the background, 
where $\mathcal{R} = {\calL}_{\rm sig}/({\calL}_{\rm sig} + {\calL}_{q \overline{q}})$ 
and ${\mathcal L}_{\rm sig}$ (${\mathcal L}_{q \overline{q}}$)
is the signal (continuum) likelihood. The variable $\mathcal R$ is then 
transformed to ${\mathcal R}^\prime \equiv \ln (\frac{\mathcal R - 0.2}{1.0 -\mathcal R})$, 
whose distribution for signal or backgrounds is easily modeled by analytical functions.

Background contributions from $\Upsilon(4S) \to B\overline B$ events are
investigated with large MC samples that include $B$ decays to final states 
with and without charm mesons. After all selection requirements are imposed,
backgrounds with charm mesons are found to be negligible; 
charmless backgrounds from multibody $B$ decays are present at negative $\Delta E$ values.  
We also identify feed-across backgrounds from other $B \to hh$ channels, 
which are typically shifted by 45 MeV in $\Delta E$ due to $K$-$\pi$ misidentification.

Signal yields are extracted by performing unbinned extended 
maximum likelihood fits to the ($M_{\rm bc}$, $\Delta E$, ${\mathcal R}^\prime$)
distributions of the candidates.
The likelihood function for each mode is 
\begin{eqnarray}
\label{fit}
 \mathcal{L} & = &  e^{ -\sum_{j} N_j}
\times \prod_i (\sum_j N_j \mathcal{P}^i_j)\;\mbox{, where}  \nonumber\\
\mathcal{P}^i_j & = &\frac{1}{2}[1- q^i \cdot {\cal A}_{CP, j}]
\mathcal{P}_j(M^i_{\rm bc}, \Delta E^i, {\mathcal R}^{\prime i})\;.
\end{eqnarray}
Here, $i$ is the event index and $N_j$ is the yield of events for the category $j$,
which indexes signal, continuum,
feed-across, and other charmless $B$ decays.
$\mathcal{P}_j(M^i_{\rm bc}, \Delta E^i, {\mathcal R}^{\prime i})$ is the PDF in
$M_{\rm bc}$, $\Delta E$, and ${\mathcal R}^{\prime}$ for the $i$-th event. 
The flavor $q$ of the $B$-meson candidate is $+1$ ($-1$) for $B^+$ and $B^0$
($B^-$ and $\overline{B}{}^0$);
${\cal A}_{CP, j}$ is the direct $CP$ asymmetry for category $j$.
For $CP$ specific modes,
$\mathcal{P}^i_j$ in Eq. (1) is simply
$\mathcal{P}_j(M_{\rm bc}^i, \Delta E^i, {\mathcal R}^{\prime i})$.
The validity of the three-dimensional fit is checked by large ensemble tests using MC events and studies of data in high statistics control samples of 
$B^{+}\to \overline{D}{}^{0}(K^{+}\pi^{-})\pi^{+}$ and $B^+\to \overline{D}{}^{0} (K^+\pi^-\pi^0)\pi^+$ decays.
The measured branching fractions for the control samples are consistent with the corresponding 
world-average values~\cite{ratios}.
Our transition to a three-dimensional fit, compared to the two-dimensional fit of previous 
publications~\cite{belleacpkpi, bellebr}, results in an effective gain in luminosity of 
32\%, 33\% and 466\%~\cite{impro23d} for $B\to K^+\pi^-$, $K^+\pi^0$ and $K^+K^-$ decays, respectively, 
as evaluated by ensemble tests.

We perform three separate simultaneous fits for pairs of modes that feed across into each other: 
(a) $B^0\to K^+\pi^-$ and $B^0\to \pi^+\pi^-$, (b) $B^+\to K^+\pi^0$ and $B^+\to \pi^+\pi^0$, and 
(c) $B^+\to K^0\pi^+$ and $B^+\to {\overline K}{}^0K^+$.  
For these fits, feed-across fractions are constrained according to the identification efficiencies and 
misidentification probabilities of charged kaons and pions.  
The $B^0\to K^+K^-$ channel is fitted alone, with the $B^0\to K^+\pi^-$ branching fraction fixed to 
the value obtained from fit (a).
The $K^0 \pi^0$ and $K^0 {\overline K}{}^0$ channels are fitted independently, 
as they have no feed-across contribution from other modes.

\begin{figure}[bh!]
\vspace{-0.3cm}
\includegraphics[width=0.44\textwidth]{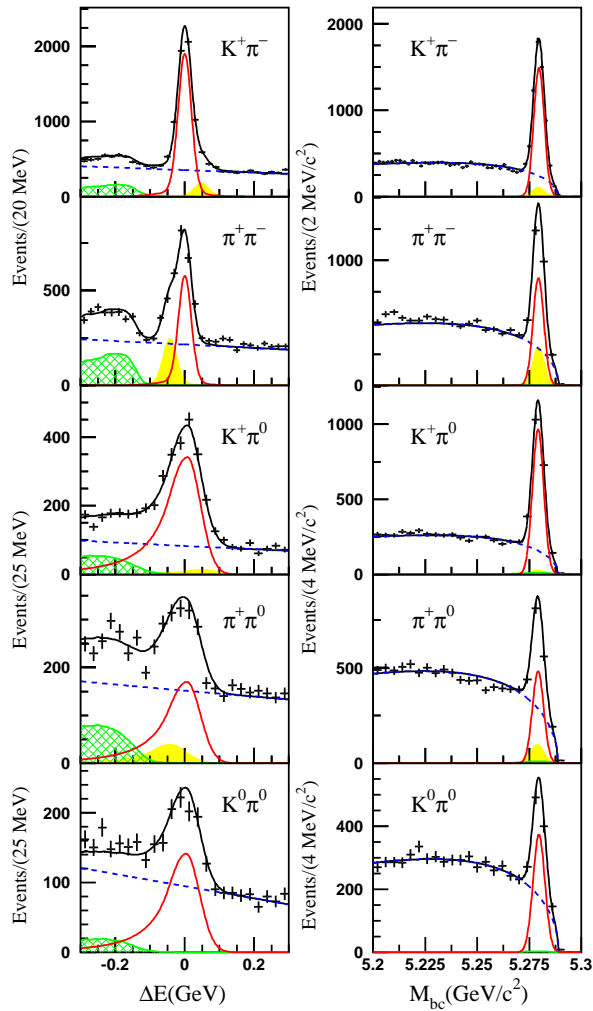}
\caption{$\Delta E$ (left) and $M_{\rm bc}$ (right) distributions for
$B^0\to K^+\pi^-$, $B^0 \to \pi^+ \pi^-$, $B^+ \to K^+ \pi^0$, $B^+ \to \pi^+ \pi^0$ and $B^0 \to K^0 \pi^0$ candidates.
Points with error bars represent the data, 
while the curves denote various components of
the fit: signal (solid red), continuum (dashed blue), charmless $B$ background
(hatched green), background from misidentification (filled yellow),
and sum of all components (solid black). The $\de$ and $\Mbc$ projections 
of the fits are for events in the $\mathcal{R}^\prime$ signal enhanced region 
($\mathcal{R}^\prime>1.47$ for $K^+\pi^-$, $\pi^+\pi^-$, and $K^0\pi^0$;
$\mathcal{R}^{\prime}>2.71$ for others) 
and ${M_{\rm bc}}>5.27$ \rm{GeV}/$c^2$ or 
$-0.14$ $(-0.06)$ \rm{GeV} $<\de<$0.06\,\rm{GeV} with (without) a $\pi^0$ in the final state.}
\label{fig:kpi}
\end{figure}

The PDFs for signal and feed-across are modeled in $M_{\rm bc}$ with a single Gaussian function, 
in $\Delta E$ with a Crystal Ball function \cite{cbf} (a double Gaussian 
function) for the modes with (without) a $\pi^0$,
and in ${\mathcal R}^\prime$ with a double or triple Gaussian function. 
Since large signals are expected for 
$B^0\to K^+ \pi^-, \pi^+\pi^-,$ and $K^0\pi^+$ decays,  
both the means and widths for $\Mbc, \de$ and ${\mathcal R}^\prime$ are floated in the fit.
For the three $h \pi^0$ modes, the $\Mbc$ means and widths are allowed to vary; 
the $\de$ means are also floated while assuming the same shift relative to 
MC values;
the ${\mathcal R}^\prime$ means and widths as well as the $\de$ widths are fixed to 
MC values after calibrating for the data-MC differences as evaluated with control samples~\cite{csde,cs}.
For the low-statistics $B^+\to {\overline K}{}^0 K^+$ decay, 
the means and widths for $\Mbc, \de$ and ${\mathcal R}^\prime$ are scaled by 
the relative positions and constant factors with respect to the parameters of $B^+ \to K^0 \pi^+$.
For the $K^0 {\overline K}{}^0$ mode, all parameters of signal PDFs 
are first fixed to MC values and then adjusted according to 
calibration factors obtained with the control sample~\cite{cs}.
In the $B^0\to K^+K^-$ fit, a triple Gaussian function is used to model the large amount of
feed-across from $B^0\to K^+\pi^-$, which includes a $\Delta E$ tail.
The means and widths for $K^+ K^-$ PDFs
are scaled by the relative values with respect to floating $K^+\pi^-$ parameters.

The continuum background PDF is described by the product of a first- or a second-order
Chebyshev polynomial for $\Delta E$, an ARGUS function~\cite{argus} for $\Mbc$, 
and a double Gaussian function for $\mathcal {R}^\prime$, modeled using off-resonance data.
The $\Delta E$ shape coefficients, the ARGUS slope parameter, and the 
${\mathcal R}^\prime$ mean and width are free parameters in the fit.
A slight correlation ($|r_{ij}|<3\%$) between the $\de$ shape coefficients and ${\mathcal R}^\prime$ 
is found in continuum events. Therefore, the continuum $\de$ shape coefficients 
are allowed to vary in four different ${\mathcal R}^\prime$ regions.  
For charmless $B$ backgrounds, a two-dimensional  
histogram is used for $(\Mbc,\de)$
to account for the correlation between these variables, while 
a double Gaussian function is employed for ${\mathcal R}^\prime$.

Projections of the fit in $M_{\rm bc}$ and $\Delta E$ are shown in 
Figs.~\ref{fig:kpi} and~\ref{fig:kk}, while projections in 
${\mathcal R}^\prime$ can be found in the appendix.
%Ref.~\cite{epaps}. 
Table \ref{tab:br} summarizes the fit results for all modes.
Assuming the production rates of $B^+B^-$ and $B^0\overline{B}{}^0$ pairs 
to be equal at the $\Upsilon(4S)$ resonance,
the branching fraction for each mode is calculated by dividing the 
fitted signal yield by the number of $B\overline{B}$ pairs and the reconstruction efficiency. 
Significant signals are observed in all channels except $B^0\to K^+K^-$ (which has $1.2\sigma$ significance).
An upper limit at 90\% confidence level on the branching fraction for this mode is obtained by integrating 
the likelihood distribution, which is convolved with a Gaussian function whose 
width equals the systematic uncertainty.

The fitting systematic uncertainties are due to signal PDF modeling, 
feed-across constraints  
and charmless $B$ background modeling. 
The PDF modeling uncertainties are estimated from the differences in signal 
yields while varying the calibration factors of signal PDFs 
by one standard deviation.
The uncertainty that arises from the modeling of final state radiation (FSR) 
is determined by lowering the photon energy threshold in PHOTOS~\cite{photos}
from 26 MeV (default) to 2.6\,MeV to derive a new set of signal PDFs, 
and subsequently a new fitted yield.
For $h^+ \pi^0$ modes, the latter uncertainties are negligible since photon shower leakage in the calorimeter 
causes a much larger $\Delta E$ tail than FSR.
Uncertainties in $K$-$\pi$ misidentification probabilities and fractions of 
feed-across events account for the dominant systematic uncertainty of 42.17\% 
in the $B^0\to K^+K^-$ channel, 
and 0.18\% to 2.28\% for the other modes.
Also a $0.45\%$ fitting bias for $B^0\to K^+\pi^-$ is incorporated by taking 
half of the yield deviation ratio
in ensemble tests with the simultaneous fit. 
The systematic uncertainties due to charmless $B$ backgrounds
 are evaluated by measuring the difference in the fitted yield 
after changing the fitting region to $\Delta E > -0.12\, {\rm GeV}$.
The above deviations in the signal yield are added in quadrature to obtain
the overall systematic error due to fitting. 

\begin{figure}[bh!]
\vspace{-0.3cm}
\includegraphics[width=0.44\textwidth]{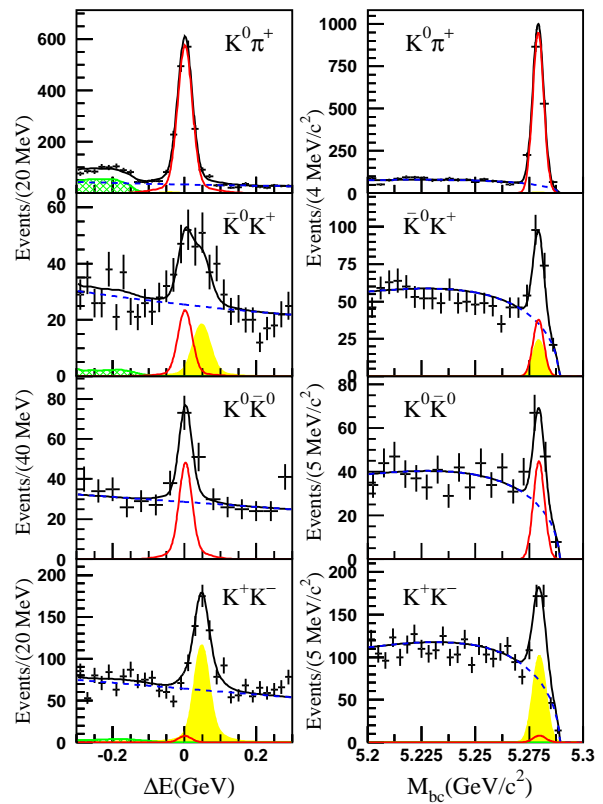}
\caption{$\Delta E$ (left) and $M_{\rm bc}$ (right) distributions for
$B^+ \to K^0\pi^+$, $B^+ \to {\overline K}{}^0 K^+$, $B^0 \to K^0{\overline K}{}^0$ 
and $B^0 \to K^+ K^-$ candidates. 
The selections for fit projections and PDF component descriptions are identical to those  
in Fig.~\ref{fig:kpi}
($\mathcal{R}^\prime>1.47$ for $K^0 {\overline K}{}^0$; $\mathcal{R}^{\prime}>2.71$ for others).}
\label{fig:kk}
\end{figure}

\begin{table*}[th]
\begin{center}
\caption{Systematic uncertainties (\%) on the measured branching fractions of $B\to hh$.}
\begin{tabular}{lcccccccccc} 
\hline\hline
				Source	& $K^+\pi^-$ & $\pi^+ \pi^-$ & $K^+ \pi^0$ & $\pi^+ \pi^0$ & ${\overline K}{}^0 K^+$ & $K^0 \pi^+$ & $K^0 {\overline K}{}^0$ & $K^0 \pi^0$ & $K^+ K^-$ \\
\hline
Tracking 				& $0.70$ & $0.70$ & $0.35$ & $0.35$ & $0.35$ & $0.35$ & - & - & $0.70$ \\
${\cal R}_{K/\pi}$     				& $1.65$ & $1.72$ & $0.78$ & $0.86$ & $0.80$ & $0.86$ & - & - & $1.58$ \\ 
$\mathcal{R}>0.2$  	& $0.55$ & $0.24$ & $0.59$ & $0.92$ & $0.91$ & $0.80$ & $0.84$ & $1.04$ & $0.28$ \\
MC statistics				& $0.16$ & $0.15$ & $0.18$ & $0.17$ & $0.20$ & $0.19$ & $0.24$ & $0.23$ & $0.16$ \\
$N_{B\overline B}$ 	& $1.37$ & $1.37$ & $1.37$ & $1.37$ & $1.37$ & $1.37$ & $1.37$ & $1.37$ & $1.37$ \\
$\pi^0$ 					& - & - & $4.0$ & $4.0$ & - & - & - & $4.0$ & - \\
$K_S^0$ 					& - & - & -     & -     & $1.68$ & $1.68$ & $3.36$ & $1.68$ & - \\
Signal PDF 				& $0.28$ & $^{+0.49}_{-0.51}$ & $0.43$ & $^{+0.89}_{-0.66}$ & $^{+0.64}_{-0.63}$ & $0.18$ & $^{+1.02}_{-1.00}$ & $1.80$ & $^{+6.76}_{-5.16}$ \\
Feed-across  	& $0.49$ & $^{+1.30}_{-1.80}$ & $0.42$ & $1.19$ & $^{+2.28}_{-2.25}$ & $0.18$ & - & - & $42.17$ \\
Fitting bias & $0.45$ & - & - & - & - & - & - & - & - \\
PHOTOS					& $1.2$ & $0.8$ & -      & -      & $0.8$ & $1.2$ & - & - & $5.0$ \\
Charmless $B$      			& $1.25$ & $1.77$ & $0.35$ & $4.53$ & $2.01$ & $0.97$ & - & $0.51$ & $1.75$ \\
\hline 
Total 						& $2.99$ & $^{+3.33}_{-3.56}$ & $4.41$ & $^{+6.51}_{-6.48}$ & $^{+4.08}_{-4.06}$ & $2.95$ & $^{+3.87}_{-3.86}$ & $5.03$ & $^{+43.09}_{-42.87}$ \\
\hline\hline
\end{tabular}
\label{systotal}
\end{center}
\end{table*}

The systematic error in efficiency caused by the likelihood ratio cut, $\mathcal{R}>0.2$, 
is investigated using control samples~\cite{cs}.
The systematic uncertainty due to charged-track reconstruction efficiency is
estimated to be 0.35\% per track using partially
reconstructed $D^{*+}\to D^0(\pi^+\pi^-\pi^0)\pi^+$ events.
The systematic uncertainty due to the ${\cal R}_{K/\pi}$ selection, which
is around 0.8\% for kaons and 0.9\% for pions, is determined from a study of 
the $D^{*+}\to D^0(K^- \pi^+)\pi^+$ sample;
the systematic uncertainties on $K_S^0$ and $\pi^0$ reconstruction are 
studied using the $D^*\to D^0 (K^0_S \pi^+\pi^-)\pi$ sample and the yield ratio
between $\eta\to\pi^0\pi^0\pi^0$ and $\eta\to \pi^+\pi^-\pi^0$, respectively.
The systematic uncertainty due to the error on the total number of $\bb$ pairs 
is 1.37\% ~\cite{nbb}. The uncertainty due to signal MC statistics is 0.2\%. 
The final systematic uncertainty is obtained by summing all these contributions in quadrature and Table ~\ref{systotal} summarizes all the systematic 
uncertainties.

\begin{figure}[h!]
\vspace{-0.3cm}
\includegraphics[width=0.44\textwidth]{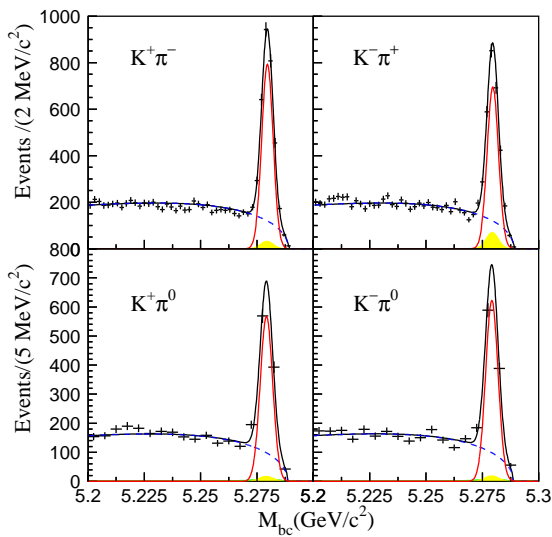}
\caption{The $M_{\rm bc}$ distributions for $B^0/{\overline B}{}^0 \to K^\pm\pi^\mp$ (top) and $B^\pm \to K^\pm \pi^0$ (bottom).   
The selections for fit projections and PDF component descriptions are identical to those described in Fig.~\ref{fig:kpi}.}
\label{fig:acpkpi}
\end{figure}

Out of the five flavor-specific decay modes presented in Table \ref{tab:br}, 
clear evidence for direct $CP$ asymmetry is found only in the $B^0\to K^+ \pi^-$ channel. 
The ${\cal A}_{CP}$ systematic errors due to fitting 
are estimated with the same procedure as applied for the branching fractions. 
Possible detector bias due to tracking acceptance and ${\cal R}_{K/\pi}$ 
selection for $h^0\pi^+$ modes 
are evaluated using the measured ${\cal A}_{CP}$ values from the continuum.
Since there is a negligible proton contamination arising from $p$-$\pi$ misidentification in the continuum, 
we conservatively assign its ${\cal A}_{CP}$ value as the systematic 
uncertainty: this  is $0.66\times 10^{-2}$ for $B^+\to \pi^+ \pi^0$ and 
$0.63\times 10^{-2}$ for $B^+ \to K^0\pi^+$. 
With regard to the detector bias for $h^0K^+$ modes, a sizable number 
of protons are included due to $p$-$K$ misidentification in continuum events.
Therefore, the possible bias is more reliably estimated using $D^+_s\to 
\phi(K^+K^-) \pi^+$ and $D^0\to K^-\pi^+$ samples \cite{acpk} and is found to 
be $(+0.33\pm 0.19)\times 10^{-2}$; 
we correct the ${\cal A}_{CP}$ values for the bias 
and assign $0.19\times10^{-2}$ as the systematic uncertainty on 
${\cal A}_{CP}$. 
For the bias of the charged kaon and pion identification in the 
$K^+\pi^-$ mode, we shift the ${\cal A}_{CP}$ value by 
$-0.33\times 10^{-2}$ and quote $0.67\times 10^{-2}$ as the systematic 
uncertainty for the residual bias. 
For ${\overline K}{}^0K^+$ and $K^0\pi^+$ modes, we shift ${\cal A}_{CP}$ 
further for the measured
$CP$ asymmetry induced by the SM $K^0-{\overline K}{}^0$ mixing: 
${\cal A}_{CP}(K^0)=(+0.332\pm0.006)\%$~\cite{ratios}.
The quadratic sum of the fitting and bias 
uncertainties gives the total ${\cal A}_{CP}$ systematic error, which ranges 
from 0.002 to 0.007.
Compared to our previous measurement of ${\cal A}_{CP}(K^+\pi^-)$ 
\cite{belleacpkpi}, the current result, ${\cal A}_{CP}(K^+\pi^-) =
 -0.069\pm 0.014 \pm 0.007$, differs by 0.025 due to a smaller measured 
central value in the newest data set of $237 \times 10^6 B{\overline B}$ pairs. 
Aside from this difference, the measurement is consistent with our previous publication 
and other experimental results \cite{babaracphh,cdfacpkpi,lhcbacpkpi}.  
Furthermore, the updated difference of $CP$ asymmetries 
$\Delta {\cal A}_{K\pi}={\cal A}_{CP}(K^+\pi^0) - 
{\cal A}_{CP}(K^+\pi^-)$  
is given by $+0.112 \pm 0.027 \pm 0.007$ 
with significance of $4.0\sigma$; this confirms our earlier result, as evident in Fig.~\ref{fig:acpkpi}.

The ratios of partial widths for $B\to K\pi$ and $B\to \pi\pi$ can be used to search for NP~\cite{buras,rn1,rn2}.
These ratios are obtained from the measurements listed in Table~\ref{tab:br}.
The ratio of charged to neutral $B$ meson lifetime, $\tau_{B^+}$/$\tau_{B^0}$ = 1.079 $\pm$ 0.007 \cite{ratios}, is used to convert 
branching fraction ratios into partial width ratios (see Table~\ref{ratio}). 
The total uncertainties are reduced because of the cancellation of common systematic uncertainties.
These ratios are compatible with SM expectations \cite{buras,pqcd,rn1,rn2} 
and supersede our previous results~\cite{bellebr}.
The partial widths and $CP$ asymmetries are used to test the violation of a sum rule~\cite{sumrule} 
given by ${\cal A}_{CP}(K^+\pi^-) + {\cal A}_{CP}(K^0\pi^+) \frac{\Gamma(K^0\pi^+)}{\Gamma(K^+\pi^-)}
- {\cal A}_{CP}(K^+\pi^0)\frac{2\Gamma(K^+\pi^0)}{\Gamma(K^+\pi^-)} - 
{\cal A}_{CP}(K^0\pi^0)\frac{2\Gamma(K^0\pi^0)}{\Gamma(K^+\pi^-)}=0$; 
 the sum is found to be $-0.270\pm 0.132\pm 0.060$ 
($1.9\sigma$ significance), using 
the results in Table \ref{tab:br}, \ref{ratio} and 
${\cal A}_{CP}(K^0\pi^0) = +0.14\pm 0.13\pm 0.06$~\cite{acpk0pi0}; 
this is still compatible with the SM prediction.
All of these results provide useful constraints to NP models 
and our uncertainties are now comparable with 
those of the corresponding theoretical calculations.

\begin{table}[h!]
\begin{center}
\caption{Partial width ratios of $B \to K \pi$ and $\pi \pi$ decays. The errors
are quoted in the same manner as in Table \ref{tab:br}.  }
\begin{tabular}{lcccccccc}\hline \hline
Modes & Ratio \cr
\hline
2$\Gamma(K^+ \pi^0)$/$\Gamma(K^0\pi^+)$ & $1.053 \pm 0.034 \pm 0.052$\cr
$\Gamma(K^+ \pi^-)$/2$\Gamma(K^0 \pi^0)$ & $1.033 \pm 0.052 \pm 0.057$ \cr
2$\Gamma(K^+\pi^0)$/$\Gamma(K^+\pi^-)$ & $1.171\pm 0.036 \pm0.055$ \cr
$\Gamma(K^+ \pi^-)$/$\Gamma(K^0\pi^+)$ & $0.899 \pm 0.026 \pm 0.030$ \cr
$\Gamma(\pi^+\pi^-)$/$\Gamma(K^+ \pi^-)$ & $0.252 \pm 0.011 \pm 0.009$ \cr
$\Gamma(\pi^+\pi^-)$/2$\Gamma(\pi^+ \pi^0)$ & $0.464 \pm 0.028 \pm 0.032$  \cr
$\Gamma(\pi^+ \pi^0)$/$\Gamma(K^0 \pi^0)$ & $0.562 \pm 0.037 \pm 0.032$ \cr
2$\Gamma(\pi^+\pi^0)$/$\Gamma(K^0\pi^+)$ & $0.490 \pm 0.024 \pm 0.033$  \cr
\hline
\hline
\end{tabular}
\label{ratio}
\end{center}
\end{table}

In conclusion, we have measured the branching fractions and direct $CP$
asymmetries for $B \to K \pi, \pi\pi$ and $KK$ decays using $772\times 10^6$ $\bb$ pairs, 
which is the final data set at Belle.
We confirm a large $\Delta {\cal A}_{K\pi}$ value with the world's smallest uncertainty.  
Including this result, the current world average is $+0.124\pm 0.022$ ($5.6\sigma$ significance)~\cite{hfag}.
We find no significant deviation from SM expectations on the partial width ratios and 
the ${\cal A}_{CP}(K\pi)$ sum rule, 
and these measurements continue to constrain the parameter space for NP.
We report a new upper limit for $B^0 \to K^+ K^-$ that is improved by a factor of two 
over the current most restrictive limit~\cite{bellebr} 
and is consistent with the latest LHCb result~\cite{lhcbacpkpi}.
Compared to previous studies,
all systematic uncertainties are decreased, including tracking, kaon/pion identification, 
$K^0_S$ reconstruction efficiencies, and the likelihood ratio requirement. 
The inclusion of the three-dimensional fit and improvements in systematic studies 
have substantially reduced the uncertainties for all channels 
and have increased the effective size of the data set.
The uncertainties for partial width ratios are all improved, especially for 
$R_c$ (by a factor of 1.6) and $R_n$ (by a factor of 1.4).

We thank the KEKB group for excellent operation of the
accelerator; the KEK cryogenics group for efficient solenoid
operations; and the KEK computer group, the NII, and 
PNNL/EMSL for valuable computing and SINET4 network support.  
We acknowledge support from MEXT, JSPS and Nagoya's TLPRC (Japan);
ARC and DIISR (Australia); NSFC (China); MSMT (Czechia);
DST (India); INFN (Italy); MEST, NRF, GSDC of KISTI, and WCU (Korea); 
MNiSW (Poland); MES and RFAAE (Russia); ARRS (Slovenia); 
SNSF (Switzerland); NSC and MOE (Taiwan); and DOE and NSF (USA).

%\newpage

\vspace{0.5cm}
\appendix
%\begin{flushleft}
\onecolumngrid

\vspace{0.5cm}
{\bf APPENDIX}

\begin{figure*}[th]
\vspace{-0.3cm}
\includegraphics[width=0.6\textwidth]{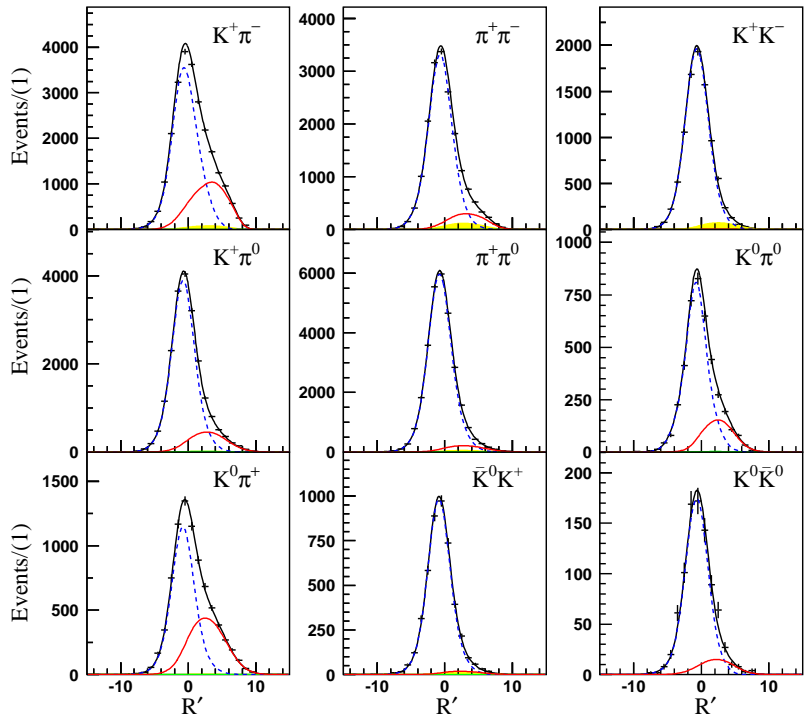}
\caption{$\mathcal{R}^\prime$ distributions for candidates of all channels.
The selections for fit and PDF component descriptions are identical to those described
in Fig.~\ref{fig:fig1}. The $\mathcal{R}^\prime$ projections of the fit are events in the
${M_{\rm bc}}>5.27$ \rm{GeV}/$c^2$ and $-0.14$ $(-0.06)\,\rm{GeV} <\de<0.06\,\rm{GeV}$
with (without) a $\pi^0$ in the final state.}
\label{fig:lr}
\end{figure*}
%\end\end{flushleft} 

\begin{thebibliography}{99}
\bibitem{conju} The inclusion of the charge-conjugate decay is implied, unless explicitly stated otherwise.
\bibitem{impro23d} 
In the previous publications, 
a tighter cut is used for the low statistics channel to suppress the high background-level.
Therefore, the luminosity gain for $K^+K^-$, which comes from the three-dimensional fit,
is substantially larger than $K^+\pi^-$ and $K^+\pi^0$ for the same ${\cal R}$ cut.
\bibitem{csde} The sample for $\de$ width calibration in $h\pi^0$ is 
$D^0$ decays to $K^-\pi^+\pi^0$ with a  
$p_{\pi^0}>$1.5 ${\rm GeV}$/$c$ requirement; the sample for $K^0 {\overline K}{}^0$ calibration 
is $D^0$ decays to $\pi^+\pi^-K^0$ with a requirement  
on $p_{K^0}>$1.0 ${\rm GeV}$/$c$.
\bibitem{cs} The control samples used for calibrating the means as well as widths for $\Mbc$ or ${\mathcal R}^\prime$, 
the means for $\de$, and for estimating systematic uncertainties due to the likelihood ratio requirement are:
(a) $B^+\to \overline{D}{}^{0} (K^+\pi^-)\pi^+$ for the $h^+ h^-$, $K^0\pi^+$, and ${\overline K}{}^0 K^+$ channels, 
(b) $B^+\to \overline{D}{}^{0} (K^+\pi^-\pi^0)\pi^+$ for the $h\pi^0$ channels, and
(c) $B^+\to \overline{D}{}^{0} (K^+\pi^- K^0)\pi^+$ for $K^0 {\overline K}{}^0$.
\bibitem{subeff} The reconstruction efficiency of $B^0\to K^0 {\overline K}{}^0$ channel 
accounts for the $K^0_S K^0_S$ decay, which corresponds to half of the $B^0\to K^0 {\overline K}{}^0$ contribution. 
The $\varepsilon \times {\cal B}_s$ term has already considered this.
\bibitem{nbb} The number of $\bb$ pairs
  is estimated based on the number of hadronic
  events measured in on-resonance data, subtracting the number of
  hadronic events in off-resonance data.  The off-resonance values
  are corrected for efficiency differences between on- and off-resonance
  configurations, as well as luminosity scaling.  After the subtraction,
  systematic uncertainty arises from the luminosity scaling, which is
  estimated using the $e^+e^- \to \mu^+ \mu^-$ events.
\end{thebibliography}
\end{document}